%% file: main.tex
\documentclass[sigconf]{acmart}
\input{package}

\AtBeginDocument{%
  }

\setcopyright{acmlicensed}
\copyrightyear{2025}
\acmYear{2025}
\acmDOI{XXXXXXX.XXXXXXX}

\acmConference[ACSAC '25]{Proceedings of the 39th Annual Computer Security Applications Conference}{December 08--12, 2025}{Waikiki, Hawaii, USA}

\acmBooktitle{ACSAC '25: Annual Computer Security Applications Conference, December 08--12, 2025, Waikiki, Hawaii, USA}
\acmISBN{978-1-4503-XXXX-X/18/06}

\begin{document}

\title{\tool: Fast Internet-wide IPv4 and IPv6 Network Scanner}


\author{Xiang Li}
\affiliation{%
  \institution{Nankai University}
  \city{Tianjin}
  \country{China}
}
\email{lixiang@nankai.edu.cn}
\authornote{Corresponding Author.}

\author{Zixuan Xie}
\affiliation{%
  \institution{Nankai University}
  \city{Tianjin}
  \country{China}
}
\email{2312585@mail.nankai.edu.cn}

\author{Lu Sun}
\affiliation{%
  \institution{Nankai University}
  \city{Tianjin}
  \country{China}
}
\email{sunlu25@mail.nankai.edu.cn}

\author{Yuqi Qiu}
\affiliation{%
  \institution{Nankai University}
  \city{Tianjin}
  \country{China}
}
\email{qiuyuqi@mail.nankai.edu.cn}

\author{Zuyao Xu}
\affiliation{%
  \institution{Nankai University}
  \city{Tianjin}
  \country{China}
}
\email{xuzuyao@mail.nankai.edu.cn}

\author{Zheli Liu}
\affiliation{%
  \institution{Nankai University}
  \city{Tianjin}
  \country{China}
}
\email{liuzheli@nankai.edu.cn}

\renewcommand{\shortauthors}{Xiang et al.}

\begin{abstract}

\tool is an open-source network scanner designed for performing fast Internet-wide IPv4 and IPv6 network research scanning.
\tool was initially developed as the research artifact of a paper published at 2021 IEEE/IFIP International Conference on Dependable Systems and Networks (DSN '21) and then made available on GitHub.
\tool is the first tool to support fast Internet-wide IPv6 network scanning in 2020.
During the last five years, \tool has made substantial impact in academia, industry, and government.
It has been referenced in 52 research papers (15 published at top-tier security venues and 11 in leading networking societies), received over 450 GitHub stars, featured in multiple news outlets, and deployed or recommended by international companies up to date.
Additionally, \tool has contributed to the implementation of RFC documents and the discovery of various vulnerabilities.
This paper provides fundamental details about \tool, its architecture, and its impact.
  
\end{abstract}





\settopmatter{printfolios=true}

\maketitle

\section{Introduction}

The IPv6 landscape has undergone significant transformation in recent years, marked by a substantial increase in the number of networks and end-hosts supporting IPv6.
For instance, the adoption rate of IPv6 among the Alexa top 1 million websites was about 2.7\% in 2012, but has risen to about 27.1\% by August 2025~\cite{w3techs2025ipv6websites}.
Similarly, while less than 1\% of Google's users accessed services via IPv6 in 2012, this figure has increased to about 46.5\% by August 2025~\cite{google2025ipv6users}.
Additionally, APNIC reports that about 33,705 Autonomous Systems (ASes) advertise IPv6 prefixes, and the number of active IPv6 BGP entries reached about 233,938 in August 2025~\cite{huston2025bgptable}.
The advent of IPv6 has introduced a vastly expanded address space, altering address allocation principles and enabling direct end-to-end Internet communication.
Specifically, end-users can now receive one or more globally addressable IPv6 prefixes from their Internet Service Providers (ISPs), shifting the address assignment paradigm from a ``single address'' model to a ``multiple prefixes'' model~\cite{iab2001iabiesg, narten2011ipv6, carpenter2015analysis}.

Given these developments, it is imperative to investigate the applicability and security issues associated with IPv6 networks.
To facilitate large-scale Internet-wide service measurements, rapid IPv4 network scanning techniques such as ZMap~\cite{durumeric2013zmap} and Masscan~\cite{masscan} have been developed.
These tools can be employed to monitor botnet activities~\cite{antonakakis2017understanding}, assess protocol deployment~\cite{alldalky2019look}, and identify vulnerabilities~\cite{aviram2016drown, li2017large, hastings2016weak}, contributing network scanning.

\noindent
\textbf{Research Problem.}
It has long been acknowledged~\cite{bellovin2006worm} that the vast address space of IPv6 networks makes exhaustive probing prohibitively expensive. Although several sophisticated techniques have been developed to identify active 128-bit end-hosts by inferring address patterns and structures~\cite{ullrich2015on, foremski2016entropy, murdock2017target, song2020towards}, and approaches such as passive collection~\cite{fiebig2017something, strowes2017bootstrapping, borgolte2018enumerating, plonka2015temporal, hu2018measuring} and hitlists~\cite{fan2010selecting, gasser2016scanning, gasser2018clusters} have been employed, these methods are significantly limited by either the diversity of seed or the complexity of the generation algorithms.
Consequently, an effective approach for global IPv6 network scanning remains elusive, posing a major challenge for studying IPv6 network security.
To overcome this obstacle, \tool was developed and open-sourced for performing fast IPv6 and IPv4 network scanning.
\textit{\tool is the first tool to provide support to fast Internet-wide IPv6 network scanning in 2020.}

\begin{figure*}[ht]
    \centering
    \includegraphics[width=0.8\linewidth]{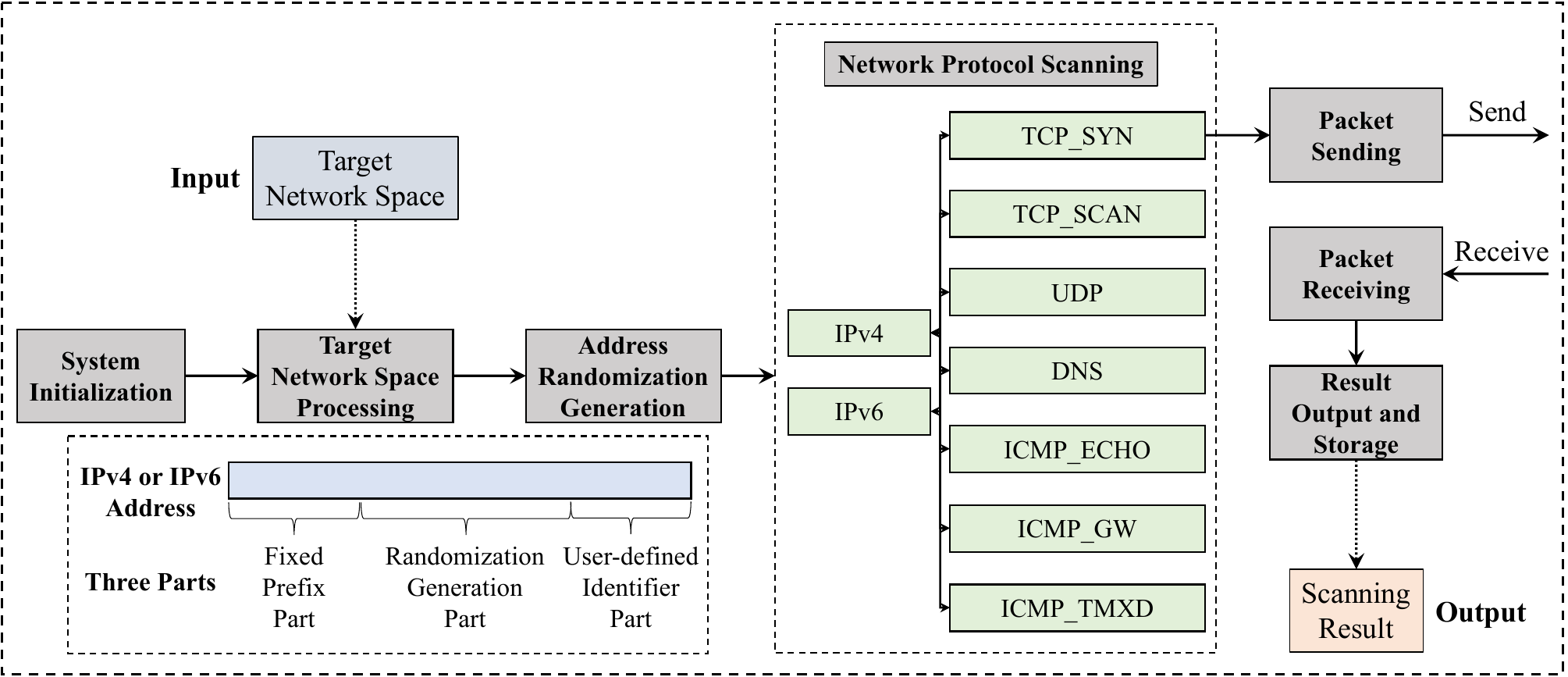}
    \caption{
    Architecture of \tool.
    }
    \Description[Architecture]{Architecture of \tool}
    \label{fig:xmap}
\end{figure*}

\noindent
\textbf{How \tool Works.}
\tool is a high-speed network scanner developed for conducting extensive IPv6 and IPv4 network research.
\textbf{It represents a comprehensive re-implementation and enhancement of ZMap}\footnote{\textit{ZMap~\cite{durumeric2024ten} mistakenly claimed XMap reused its code, which we have already refuted.}}, maintaining full compatibility with its predecessor while introducing advanced scanning techniques and achieving a probing speed of only ``5 minutes''.
\tool is capable of scanning the entire 32-bit address space in under 45 minutes.
With a 10 GigE connection and PF\_RING~\cite{pfring}, \tool can complete this scan in less than 5 minutes.
Additionally, \tool utilizes innovative IPv6 scanning methodologies to rapidly identify the IPv6 network periphery.
\tool provides unique flexibility in scanning, distinguishing itself as the only tool with this capability.
\textit{It enables random scans of network spaces of arbitrary length and position, such as 2001:db8::/32-64 and 192.168.0.1/16-20.}
\textit{It supports simultaneous probing of multiple ports.}
\tool is compatible with GNU/Linux, macOS, and BSD operating systems and includes multiple probing modules for ICMP Echo scans, TCP SYN scans, UDP probes, and DNS scans (including newly-developed stateless, stateful, or address-spoofing variants).
\tool introduces novel ICMP and DNS probing modules, which are implemented for the first time and utilized across various research scenarios.

\noindent
\textbf{Academic Impact.}
\tool is an open-source IPv4 and IPv6 network scanner that was released under the Apache License as the artifact accompanying the paper~\cite{li2021fast} published at 2021 IEEE/IFIP International Conference on Dependable Systems and Networks (DSN '21~\cite{dsn21}).
Since its release, \tool and its paper have been cited extensively in the academic literature, supporting various research projects.
Notably, 52 academic papers have referenced \tool, including 15 published at top-tier security venues.
These papers span several fields, including IPv6 network scanning~\cite{steger2023target, hou20236scan, pan2023your, zhang20246rover, williams20246sense, wang2023knocking, dahlmanns2024unconsidered, liu20236former, hu2023dscan6, yang2024efficient}, security or privacy evaluation and measurement~\cite{saidi2022one, zhang2023wolf, durumeric2024ten, chen2024zbanner, maier2023in, saidi2022characterizing, zhang2023under, zhang2024rethinking}, vulnerability discovery~\cite{li2024tudoor, olson2023doomed, xu2023tsuking, li2023maginot, li2023ghost, zhang2024resolverfuzz, li2024dnsbomb, liu2024understanding, wang2024breakspf}, defense mechanisms~\cite{he2022design, Olson2024where}, and surveys~\cite{liu2022survey, ageel2024survey, vaishnav2023from}.
Among these, one paper~\cite{pan2023your} received the Distinguished Paper Award at the 30th Annual Network and Distributed System Security Symposium 2023 (NDSS '23).
Another IEEE S\&P 2024 paper~\cite{li2024tudoor} received the 2024 Pwnie Award Nominations for Most Innovative Research~\cite{tudoorpwnie}.
They are all top-tier security venues.

\noindent
\textbf{Open-Source Artifact.}
The primary project is hosted on GitHub at \href{https://github.com/idealeer/xmap}{https://github.com/idealeer/xmap} and is actively maintained by us.
The usage documentation of \tool can be found at \url{https://github.com/idealeer/xmap/blob/master/src/xmap.1.ronn}.
Usage examples are shown at the Wiki page \url{https://github.com/idealeer/xmap/wiki}.
The tool is continuously enhanced with new features and updated to ensure ongoing compatibility.
At the time of writing, \tool comprises approximately 140k lines of C code and 68k lines of C++ code, excluding blank and comment lines.
In addition to its academic impact, as evidenced by \textit{52 citations}, \tool has received significant recognition within the open-source community, reflected in its \textit{456 stars}, \textit{74 forks}, and \textit{19 watchers}.
Although \tool has evolved significantly since its initial release as a research prototype designed to address IPv6 scanning challenges, it remains an active project with ample opportunities for further development.
Our dedication to advancing \tool is demonstrated through \textit{27 support issues}, \textit{15 release versions}, and \textit{362 commits} to date, all of which contribute to its continuous improvement and robustness.
\tool is packaged as a free Docker image~\cite{xmapdocker}, which has been downloaded \textit{595 times}, and as a Linux package~\cite{xmapgit}, which has garnered 2 votes.
\tool is reported on and recommended by various news outlets and leading companies (e.g., \cite{pentestertv, xmapnews, xmapgoogle, xmapcisco}).
\tool also contributes to the implementation of RFC documents~\cite{singh2013basic}.

Section~\ref{sec:arch} describes \tool's architecture, while Section~\ref{sec:impa} details its broad impact across academia, industry, and government fields.

\section{Architecture of \tool}
\label{sec:arch}

\tool is a new system designed for rapid probing of IPv4 and IPv6 network spaces, utilizing asynchronous decoupling and address randomization generation techniques.
It features a modular architecture that incorporates a customized network protocol scanning module, address identifier generation module, and data output and storage module.
\tool efficiently handles packet generation, sending, receiving, and comprehensive address generation while minimizing impact on target networks.
By employing asynchronous scanning and full permutation randomization, \tool effectively addresses the challenges of probing any target address, port, or network protocol, achieving swift and efficient network space probing, as well as ease of use and adaptability for further customization.

As shown in Figure~\ref{fig:xmap}, \tool includes the following modules: system initialization, target network space processing, address randomization generation, network protocol scanning, packet sending and receiving, and result output and storage.

\begin{itemize}

    \item \textbf{System Initialization Module:}
    Responsible for starting the entire probing system and initializing the parameters and files required for the probe.
    \tool automatically identifies local network interfaces, IP addresses, and MAC address lists, configures probing speed, manages blacklists and whitelists, and initializes subsequent modules.

    \item \textbf{Target Network Space Processing Module:}
    Handles the input provided by the user to determine the target probing network space.
    \tool allows flexible definition of custom scanning address ranges based on user input, providing significant versatility than ZMap~\cite{durumeric2013zmap} and Masscan~\cite{masscan}.
    The address in \tool is divided into three components: the fixed prefix, the randomization generation part, and the user-defined identifier. 
    Unlike existing systems~\cite{durumeric2013zmap, masscan} that only specify address prefixes or limited portions of address space, \tool supports scanning any segment of the IPv4 or IPv6 address space (the randomization generation part), such as from the 16th to the 24th bit (from the 32nd to the 48th bit).

    \item \textbf{Address Randomization Generation Module:}
    Randomizes target probing addresses within the input address space.
    Utilizing a permutation modulo multiplication algorithm, \tool traverses the specified address space and converts IPv4 or IPv6 addresses into custom large integers with GMP~\cite{gmp}.
    Mathematical operations are then performed to randomize address generation, distributing probing traffic across various target networks and minimizing impact.
    The completeness of address randomization is ensured through rigorous mathematical validation.
    For address spaces not designated for randomization, \tool supports address identifier generation through user specification, pattern generation, or randomization for the user-defined identifier part.

    \item \textbf{Network Protocol Scanning Module:}
    Integrates various user-defined network protocol scanning functions.
    \tool implements scanning capabilities for nearly all major network protocols, including IPv4, IPv6, TCP, UDP, and ICMP.
    It supports parallel scanning of multiple ports, discovery of network critical devices such as periphery devices, and advanced probing for certain protocols like DNS.
    Specifically for DNS, \tool is capable of sending packets with spoofed addresses, conducting stateful scans, and detecting DNS software versions.
    The module also incorporates vulnerability scanning for issues like routing loops, greatly facilitating the integration of subsequent scanning modules.
    Users have the capability to develop new scanning modules or extend existing ones to tailor probing tasks to their specific needs.

    \item \textbf{Packet Sending and Receiving Module:}
    Employs high-speed, asynchronous decoupled packet sending and receiving.
    \tool utilizes existing packet separation technologies to redesign and implement IPv4 and IPv6 packet sending and receiving interfaces.
    By integrating IPv4 and IPv6 sockets with multiple operating systems, and utilizing specific packet fields from the network protocol scanning module, \tool achieves high-speed separation and matching of IPv4 and IPv6 packets.

    \item \textbf{Result Output and Storage Module:}
    Provides multiple result output based on files or databases.
    \tool supports selective output and storage of all or specific fields of the scanning results, such as addresses, ports, and payloads, using various file formats (TXT, CSV, and DB).
    
\end{itemize}

\tool addresses the challenges of probing IPv4 and IPv6 network spaces from a novel perspective by employing an asynchronous scanning mechanism with decoupled packet sending and receiving.
Utilizing a modular approach, \tool integrates network protocol scanning, address identifier generation, and data output and storage modules to efficiently probe any target address, port, or network protocol in IPv4 and IPv6 spaces.
This approach minimizes the impact on target networks and enables rapid network space probing through asynchronous decoupling and address randomization.

\section{Impact}
\label{sec:impa}

Since its public release in April 2021, \tool has made a substantial impact across the academia, industry, and government sectors.
The development of \tool began in August 2020 (before the paper submission DDL of the DSN 2021 conference~\cite{dsn21}), and as of August 2025, it celebrates the fourth anniversary.

\subsection{Academic Impact}

In the academic domain, as the first tool supporting fast Internet-wide IPv6 network scanning, \tool and its paper were highly commended by reviewers, who noted that ``\textit{The resulting tool will be very useful as a building block in the Internet measurement and security community}''.
As anticipated, they have been extensively cited, contributing to various researches.

\begin{table}[htbp]
  \centering
  \small
  \caption{Citation Sources by Societies.}
    \begin{tabular}{lr}
    \toprule
    \textbf{Publication Societies} & \multicolumn{1}{l}{\textbf{Citations}} \\
    \midrule
    \rowcolor[rgb]{ .91,  .91,  .91} USENIX Security & 5 \\
    NDSS  & 5 \\
    \rowcolor[rgb]{ .91,  .91,  .91} ACM IMC & 4 \\
    IEEE S\&P & 3 \\
    \rowcolor[rgb]{ .91,  .91,  .91} ACM CCS & 2 \\
    IEEE ISCC & 2 \\
    \rowcolor[rgb]{ .91,  .91,  .91} IEEE/ACM ToN & 2 \\
    IEEE ICNP & 2 \\
    \rowcolor[rgb]{ .91,  .91,  .91} IEEE INFOCOM & 2 \\
    PhD Thesis & 2 \\
    \rowcolor[rgb]{ .91,  .91,  .91} ACM Computing Surveys & 1 \\
    ACM SIGCOMM CCR & 1 \\
    \rowcolor[rgb]{ .91,  .91,  .91} IEEE TMA & 1 \\
    IEEE IoT Journal & 1 \\
    \rowcolor[rgb]{ .91,  .91,  .91} Others & 19 \\
    \midrule
    Total & 52 \\
    \bottomrule
    \end{tabular}%
  \label{tab:cita}%
\end{table}%

Table~\ref{tab:cita} presents the publication societies (conference and journal) that have cited \tool or its associated paper.
These publication societies are among the most prominent and active in the fields of network security, measurement, and privacy evaluation.
In total, \tool is referenced in 52 academic publications, including 15 papers in top-tier security venues: IEEE S\&P (3), USENIX Security (5), ACM CCS (2), and NDSS (5).
Additionally, 11 papers are published in leading networking or measurement venues and journals: ACM SIGCOMM Computer Communication Review (1), IEEE/ACM ToN (2), ACM IMC (4), IEEE INFOCOM (2), and IEEE ICNP (2).

These papers span multiple meaningful research areas: 

\begin{itemize}

    \item \tool serves as a foundational tool in the comparison of IPv6 network scanning techniques~\cite{steger2023target, hou20236scan, pan2023your, zhang20246rover, williams20246sense, wang2023knocking, dahlmanns2024unconsidered, liu20236former, hu2023dscan6, yang2024efficient}, demonstrating superior speed compared to previous IPv6 address discovery methods.

    \item \tool has been employed for security and privacy evaluations, including the assessment of potential malicious services~\cite{saidi2022one, zhang2023wolf, durumeric2024ten, chen2024zbanner, maier2023in, saidi2022characterizing, zhang2023under, zhang2024rethinking}.
    It exhibits both rapid probing capabilities and broad adaptability across various network scenarios.

    \item \tool is utilized for vulnerability discovery, specifically targeting DNS cache poisoning, DoS, and email spoofing attacks~\cite{li2024tudoor, olson2023doomed, xu2023tsuking, li2023maginot, li2023ghost, zhang2024resolverfuzz, li2024dnsbomb, liu2024understanding, wang2024breakspf}.
    It effectively collects DNS record data, facilitating thorough vulnerability analysis.
    
    \item \tool and its associated techniques have inspired new defense mechanisms and contributed to the development of novel authentication schemes~\cite{he2022design, Olson2024where}.

    \item \tool is also integrated into network scanning and security survey works~\cite{liu2022survey, ageel2024survey, vaishnav2023from}, marking a significant advancement in academic research.

\end{itemize}

Among these paper above, the paper~\cite{pan2023your} titled ``Your Router is My Prober: Measuring IPv6 Networks via ICMP Rate Limiting Side Channels'' received the \textit{Distinguished Paper Award} at the 30th Annual Network and Distributed System Security Symposium 2023 (NDSS '23), a top-tier security venue.
By enhancing \tool's ICMPv6 probing technique, this paper introduces a novel measurement technique called IVANTAGE, which leverages ICMP rate limiting side channels to assess remote IPv6 networks.
IVANTAGE utilizes these side channels to ``send'' and ``receive'' packets on remote Internet nodes, effectively enabling the use of routers in distant networks as indirect ``vantage points'' without requiring direct control over them.
\tool inspired the development of this new method.
Another paper~\cite{li2024tudoor} titled ``TuDoor Attack: Systematically Exploring and Exploiting Logic Vulnerabilities in DNS Response Pre-processing with Malformed Packets'' (published at the 2024 IEEE Symposium on Security and Privacy, IEEE S\&P '24) received the 2024 Pwnie Award Nominations for Most Innovative Research~\cite{tudoorpwnie}.

\subsection{Industrial Impact}

\tool has made notable impacts in the industry:

\begin{enumerate}

    \item The tool has garnered substantial recognition within the open-source GitHub community~\cite{xmapgit}, evidenced by its 456 stars, 74 forks, and 19 watchers.
    Despite its evolution from an initial research prototype designed to tackle IPv6 scanning issues, \tool remains an active project with significant potential for further development.
    Our commitment to enhancing \tool is reflected in the resolution of 27 support issues, the release of 15 versions, and 362 commits to date, contributing to its ongoing improvement and robustness.

    \item \tool has been distributed as a free online Docker image~\cite{xmapdocker} (which has been downloaded 595 times) and a public Linux package~\cite{xmaplinux} (which has received 2 votes).
    Additionally, \tool has been recognized as an impressive scanning tool within the famous Kali Linux Hacker tool community~\cite{xmapkali} and the hacker tools website~\cite{xmapkitploit}, and collected into the awesome hacking list~\cite{xmapawesome}.
    
    \item \tool is employed by various security companies as part of their commercial products for long-term network device probing, including Qianxin~\cite{qianxin} and WebRay~\cite{webray}, and ISPs such as China Unicom, Mobile, and Telecom.
    \tool is also recommended by heads or engineers from several international companies, such as Google~\cite{xmapgoogle}, Cisco~\cite{xmapcisco}, Henkel~\cite{xmaphenkel}, National Cyber Security Services~\cite{xmapncyber}, senhasegura~\cite{xmapsan}, Ing. de Telecomunicaciones~\cite{xmapapo}, and Diyako Secure Bow~\cite{xmapdiyako}.

    \item \tool has contributed to the implementation of RFC documents, such as RFC 7084~\cite{singh2013basic} that recommends ``\textit{Any packet received by the CE router with a destination address in the prefix(es) delegated to the CE router but not in the set of prefixes assigned by the CE router to the LAN must be dropped}'', to avoid the routing loop vulnerability.
    
    \item \tool has been awarded a China patent for its rapid IPv6 network periphery discovery technique~\cite{xmappatent1}.
    Other two China patents that use \tool or describe the architecture~\cite{xmappatent2, xmappatent3} have been published and is currently under review.
    
    \item \tool has been highlighted by international media, such as Pentester Academy TV~\cite{pentestertv}, Security Online News~\cite{xmapnews}, and Hacking \& Pentest Videos~\cite{xmaphacking}, and featured in the famous 2021 West Lake cybersecurity conference's cyberspace security tools presentation~\cite{westlake}.

    \item \tool helps detect and address vulnerabilities for numerous leading global Internet companies, such as Google, Apple, Microsoft, Cloudflare, Akamai, Cisco, Oracle, Verisign, ByteDance, Ali, Baidu, Tencent, Huawei, ASUS, Netgear, Linksys and others~\cite{li2021fast, li2023ghost, li2023maginot, xu2023tsuking, li2024dnsbomb}, affecting over 6 million devices in the wild and receiving over 250 CVE or CNVD numbers.
    These vulnerabilities stem from deficiencies in protocol design and impact all software implementations.

\end{enumerate}

\subsection{Governmental Impact}

\tool has had a notable impact on government-related departments, with its IPv6 scanning technique being utilized by the China Education and Research Network (CERNET)~\cite{cernet} and People's Public Security University of China (PPSUC)~\cite{gongan}.
\tool is executed periodically by these institutions.
\tool received both the first and third prizes in the 2022/2024 IPv6 Technology Application Innovation Competition~\cite{ipv6competition} hosted by CERNET.
In 2023, \tool is also integrated into an IPv6 security demonstration platform for guaranteeing campus network security in collaboration with CERNET~\cite{ipv6tech}.
\tool and its rapid IPv6 network periphery discovery technique were featured at the 2023 World Internet Conference for IPv6~\cite{ipv6internet}.
In 2025, projects powered by XMap won the first and third prizes in the National College Student Information Security Contest, as well as Outstanding Instructor Award for advisors.

\section{Conclusion}

\tool represents a significant advancement in network scanning technology, akin to how high-speed photography transformed the study of rapid physical phenomena.
As an open-source tool, \tool has not only facilitated extensive research in IPv4 and IPv6 network scanning but has also made a profound impact across academia, industry, and government sectors.




\balance
\bibliographystyle{ACM-Reference-Format}
\bibliography{reference}

\appendix

\end{document}

%% file: package.tex
\usepackage{xspace}
\usepackage{colortbl}

\newcommand{\tool}{{\textsc{XMap}}\xspace}